\documentclass[11pt,a4paper]{article}
\usepackage[T1]{fontenc}
\usepackage{lmodern}
\usepackage[margin=2.5cm]{geometry}
\usepackage{graphicx}
\usepackage{amsmath,amssymb,bm}
\usepackage{booktabs}
\usepackage{float}
\usepackage{xcolor}
\usepackage{notoccite}
\usepackage{cite}
\usepackage{microtype}
\usepackage[font=small,labelfont=bf]{caption}
\usepackage{authblk}

\usepackage{hyperref}
\hypersetup{colorlinks=true,linkcolor=blue!50!black,citecolor=blue!50!black,urlcolor=blue!50!black}

\graphicspath{{./}}

\title{Performance Analysis of a Dual-Polarised Bowtie Antenna in the Low RF Band:\\
A Low-Cost Undergraduate Laboratory Experiment}

\author[1]{Robin Sen\thanks{\texttt{robinsen2217@gmail.com}, ORCID: \href{https://orcid.org/0009-0004-1343-4745}{0009-0004-1343-4745}}}
\author[1]{Akshay Bhatia Neriyanuri\thanks{\texttt{akshay.jeremy@gmail.com}, ORCID: \href{https://orcid.org/0009-0000-8298-9569}{0009-0000-8298-9569}}}
\author[1]{Geetanjali Sethi\thanks{Corresponding author: \texttt{getsethi@ststephens.edu}, ORCID: \href{https://orcid.org/0000-0002-7542-2748}{0000-0002-7542-2748}}}
\affil[1]{Department of Physics, St.~Stephen's College, University of Delhi, Sudhir Bose Marg, Delhi 110007, India}
\date{}

\begin{document}
\maketitle

\begin{abstract}
We have set up a simple experiment to characterise a dual-polarised bowtie dipole antenna at St.~Stephen’s College as part of  undergraduate radio astronomy experiments. The antenna provides two orthogonal linear polarisation responses, allowing us to investigate the polarisation properties of radio-frequency signals using a simple and affordable experimental setup. In this article, we present our analysis of  the performance of the antenna and demonstrate its use through polarisation-rotation measurements. This antenna is part of  the Sky Watch Array Network (SWAN) and this analysis helps us understand the response of each individual array element in the setup.
\end{abstract}

\section{Introduction}
Astronomical sources emit electromagnetic radiation across a broad range of frequencies, with different frequency ranges providing access to different astronomical phenomena \cite{condon2016essential}. Radio astronomy explores the Universe through its radio-frequency emission and has become an important complement to observations at optical and other wavelengths \cite{condon2016essential,wilson2009tools}.

Radio waves from astronomical sources are detected using antennas that work in the radio frequency (RF) band, and form the interface between the incoming radiation and the receiver system \cite{wilson2009tools}. Along with the strength and frequency of the received signal, its polarisation can also carry information about the source and the environment through which the radiation travels \cite{wilson2009tools,born1999principles}. The study of polarisation is therefore an important aspect of radio observations and provides an additional perspective on the nature of astronomical radio emission.
In this article we have used parts of an array network (SWAN) to perform simple experiments to study polarisation of radio waves. It is usually not a simple task to perform experiments to understand the very important idea of polarisation. The advantage of our experiment is that it is low cost and simple to perform. The data obtained are also easy to analyse, and the analysis can be done in a classroom.
\section{Dual-Polarised Bowtie Antenna}
Antennas are used to transmit and receive electromagnetic waves and form an essential part of wireless communication, radar, imaging, and radio astronomy systems. Different types of antennas, including dipoles, monopoles, loops, arrays, microstrip antennas, horns, and reflector antennas, are used depending on the required frequency range and radiation characteristics~\cite{balanis2005antenna,roederer2014ieee}. Their performance is commonly described by parameters such as gain, directivity, radiation pattern, bandwidth, efficiency, and polarisation~\cite{balanis2005antenna,roederer2014ieee,balanis2008fundamental}. 

Gain and directivity describe the concentration of radiated or received energy in a particular direction, while the radiation pattern describes how this energy is distributed spatially. The beamwidth of an antenna is closely related to its directivity and determines the angular extent of the main beam; a narrower beam generally corresponds to higher directivity and gain~\cite{roederer2014ieee,balanis2008fundamental}. In applications such as radar, imaging, and radio astronomy, the beamwidth and antenna aperture consequently play an important role in determining angular resolution, while a wider beam provides a larger field of view. Bandwidth is also an important consideration, particularly for systems that require operation over a wide frequency range~\cite{balanis2005antenna,balanis2008fundamental}.

Bowtie antennas offer a simple planar physical structure and stable performance over a wide bandwidth~\cite{zhang2015integrated}. These have wide applications such as wireless communications~\cite{madhav2011liquid}, terahertz wave near-field imaging~\cite{ishihara2006terahertz} and flexible RF devices~\cite{durgun2011design}. The dual-polarised bowtie antenna used in the Sky Watch Array Network (SWAN) system at St.~Stephen's College, Delhi is a Murchison Widefield Array (MWA) type antenna \cite{lonsdale2009murchison,tingay2013murchison} with a crossed pair of bowtie-shaped elements made of aluminium having a wingspan of 0.74 m and a height of 0.40 m (measured). The vertical bowtie pairs are orthogonal to each other providing dual-linear polarisation. The antenna is designed for the 80--320~MHz range. Each of the four bowtie arms is made up of four pieces of aluminium channels and an aluminium key. The keys slide into slots in the central hub, made of UV-resistant white PVC. This hub stores the two Low Noise Amplifier (LNA) boards (one for each polarisation) which are stacked vertically one above the other inside. The bottom tips are supported by dielectric feet made up of UV-resistant white PVC and having a height of 0.1~m (measured).\par

\begin{figure}[!t]
\centering
\caption{Image of dual-polarised bowtie antenna set up at St.~Stephen's College, Delhi}
\includegraphics[scale=0.22]{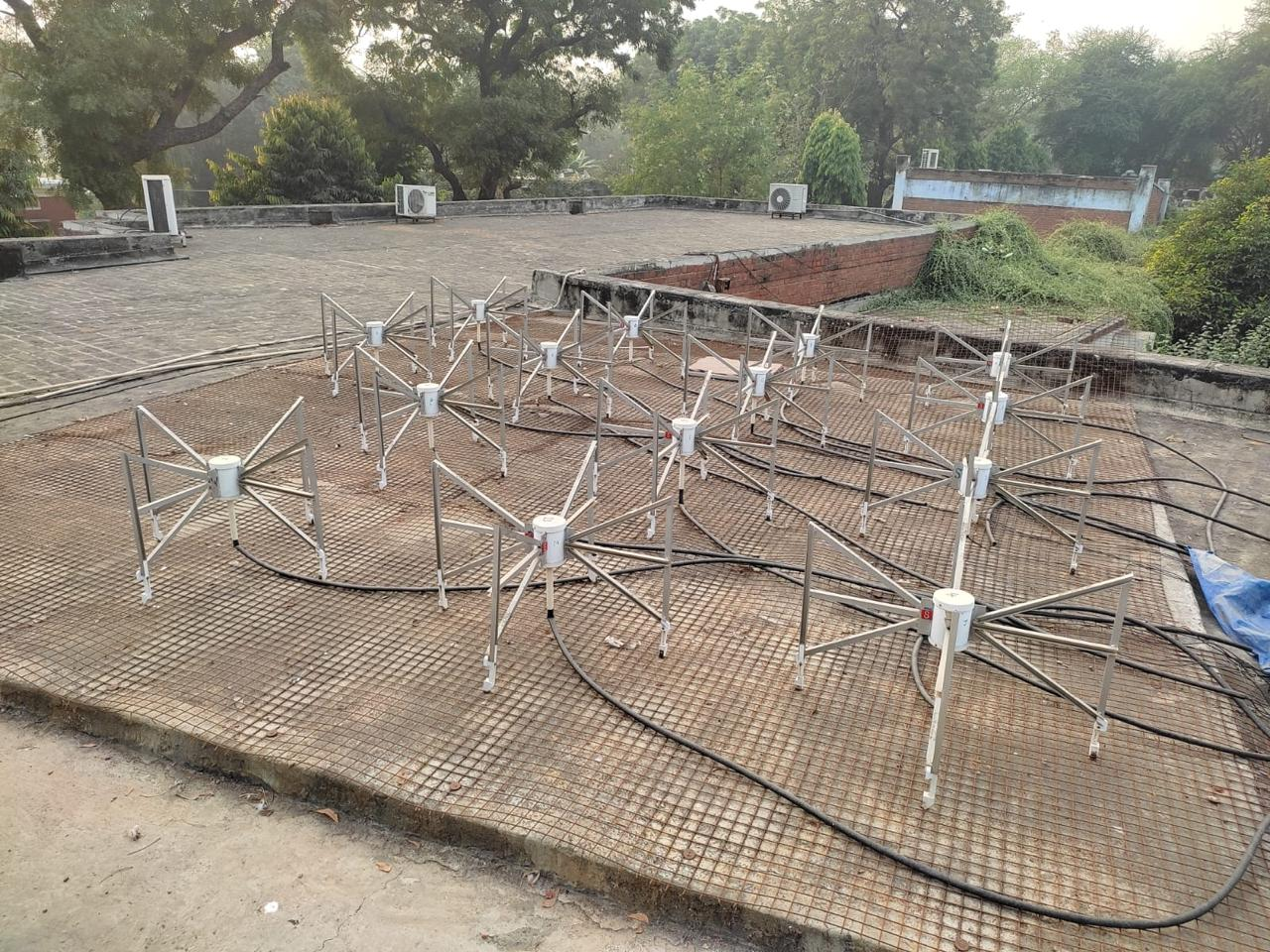}
\end{figure}

\subsection*{Radiation Pattern} 
The radiation pattern of an antenna describes the variation of its radiated power with direction and is used to determine how electromagnetic energy is distributed in space. It is commonly represented in terms of the principal E- and H-planes, with parameters such as beamwidth, main-lobe direction, and sidelobe level used to characterise the pattern \cite{balanis2015antenna,roederer2014ieee}. For a conventional half-wave dipole, the radiation pattern is approximately doughnut-shaped in three dimensions, with maximum radiation in directions perpendicular to the antenna axis and nulls along its axis. A bowtie antenna exhibits a similar dipole-like radiation behaviour, with its principal radiation directed perpendicular to the antenna axis and reduced radiation along the axis \cite{compton1987bow,zhang2015integrated}. However, unlike an ideal dipole, the radiation pattern of a bowtie can vary with frequency due to its broadband nature and finite dimensions. Experimental studies have shown that bowtie antennas generally exhibit broad, dipole-like far-field radiation patterns, with changes in beamwidth and gain occurring across the operating frequency range \cite{compton1987bow,zhang2015integrated,de2022bow}.

\begin{figure}[H]
	\centering
	\caption{Simulated radiation pattern of the dual-polarised bowtie over a perfect
		ground plane at 200~MHz, made with 4nec2 software~\cite{burke1981numerical}. (a)~Side view and (b)~top view
		of the model. (c),~(d)~3D total-gain patterns with polarisation 1 and 2
		excited, respectively. (e)~Horizontal-plane cut at elevation $=10^\circ$
		($\theta=80^\circ$), with total gain normalised to the peak of each curve.
		Blue: polarisation 1; Red: polarisation 2.}\label{rad_pattern}
	\includegraphics[width=\columnwidth]{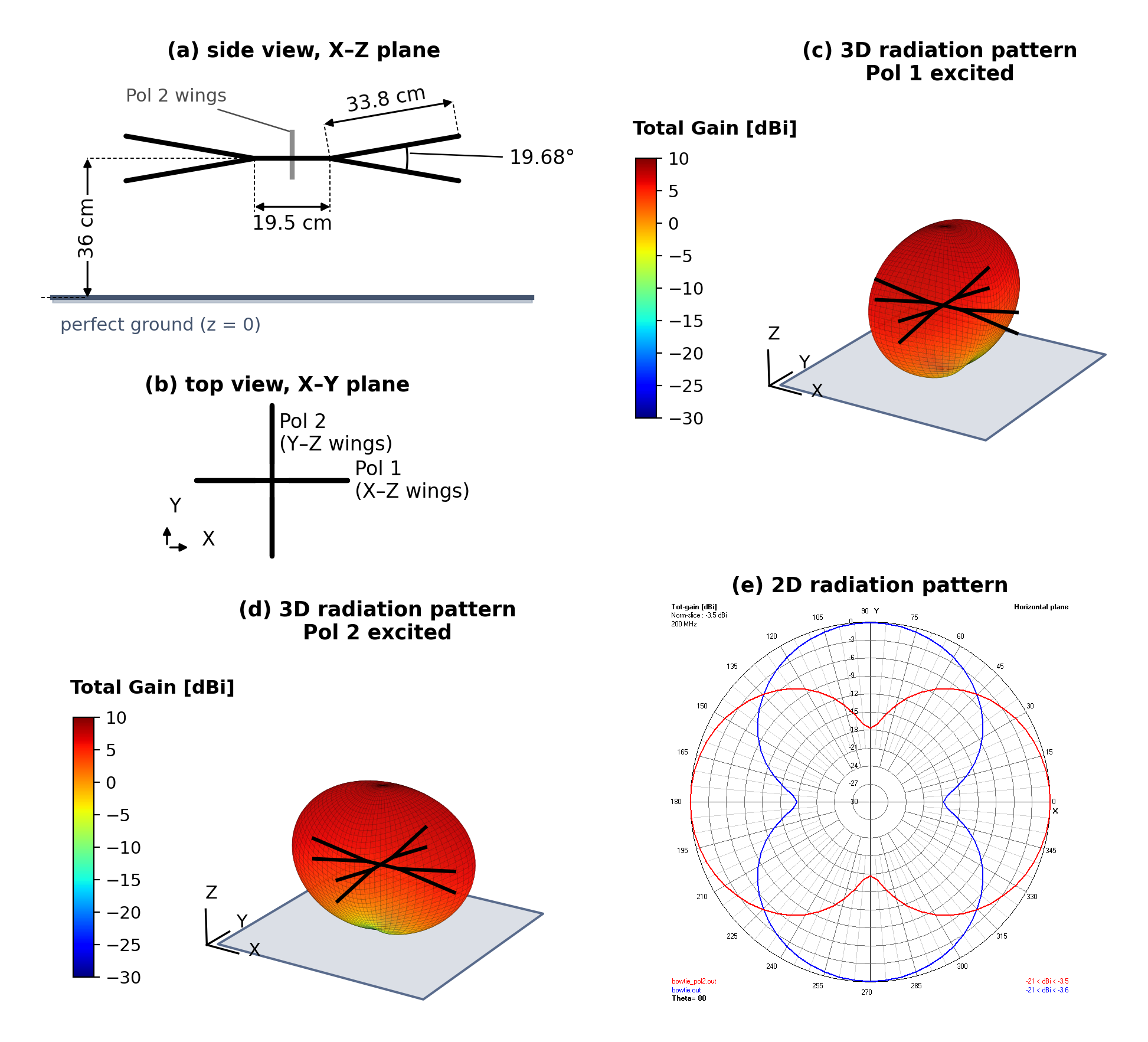}
\end{figure}

Figure~\ref{rad_pattern} shows the simulated response of the two polarisations.
In the model (Fig.~\ref{rad_pattern}(a,b)), each polarisation consists of two
wings of length 33.8~cm, with a 19.5~cm wide central region and the wings opened
by $19.68^\circ$. The feed is 36~cm above a perfect ground plane ($z=0$).
Polarisation 1 uses the wings in the $X$--$Z$ plane and polarisation 2 the wings
in the $Y$--$Z$ plane. Exciting either polarisation produces a single broad main
lobe towards the zenith, with a peak total gain of about 11.6~dBi for
polarisation 1 (Fig.~\ref{rad_pattern}(c,d)). Since a real ground screen is finite
and lossy, the gain of the deployed antenna will be somewhat lower than this
idealised value. The two 3D patterns have the same
shape, rotated by $90^\circ$ about the vertical axis, as expected from the
orthogonal arrangement of the two bowties.

The horizontal cut at $10^\circ$ elevation (Fig.~\ref{rad_pattern}(e)) shows the
main lobes of the two ports crossed, along $Y$ for polarisation 1 and along $X$
for polarisation 2. The nulls lie about 17.6~dB below the peak. The pattern
depends on the direction of arrival, and the figure-of-eight shape becomes more
prominent as the minima fall deeper when the elevation is further reduced. We show
a cut at $10^\circ$ elevation because terrestrial radio-frequency interference,
such as FM broadcast signals, arrives close to the horizon, where the responses
of the two ports differ most.
 
\subsection*{Polarisation}
Polarisation describes the orientation and time variation of the electric-field vector of an electromagnetic wave and is an important characteristic of an antenna that radiates the wave. A wave is said to be linearly polarised when the electric-field vector maintains a fixed direction as the wave propagates, while circular and elliptical polarisation occur when the field direction rotates with time \cite{balanis2015antenna,roederer2014ieee}. For a dipole antenna, the radiated wave is predominantly linearly polarised, with the electric field oriented parallel to the antenna element. A bowtie antenna similarly produces predominantly linear polarisation, with its polarisation direction determined by the orientation of the bowtie element \cite{compton1987bow}. In the Murchison Widefield Array (MWA), two bowtie elements are arranged orthogonally, providing two independent, approximately orthogonal linear polarisation components. This dual-polarised configuration allows the two components of the incoming electromagnetic field to be measured independently \cite{tingay2013murchison,lonsdale2009murchison}. To specify how well an antenna distinguishes its two polarisations, the cross-polarisation discrimination (XPD) can be measured~\cite{8760360}. It is the ratio of power measured by one state of polarisation to that received in the orthogonal state, when the source polarisation is aligned with the former. Thus, a higher XPD means less cross-talk~\cite{jaiswal2025study}.

\subsection*{SWAN}
The Sky Watch Array Network (SWAN) is a low-frequency radio astronomy initiative aimed at developing a distributed array of radio observing stations for studying transient radio sources and performing high-angular-resolution observations. By combining observations from stations separated by large distances, SWAN can exploit interferometric techniques to achieve higher angular resolution than can be obtained with a single observing station. The system incorporates wide-band receiver instrumentation to support observations over a broad frequency range, with the receiver development focused on providing a practical and scalable platform for low-frequency radio astronomy. The SWAN project also emphasises the development of cost-effective instrumentation that can facilitate the deployment of radio observing systems at multiple locations \cite{rao2021wide}. SWAN uses dual-polarised bowtie antennas for its observations. A SWAN setup has also been established at St.~Stephen’s College, University of Delhi, as part of efforts to promote hands-on training and low-cost radio astronomy instrumentation. However, the RFI due to FM stations in Delhi lies within the operational frequency range of SWAN and thus restricts the observation time to night, when the RFI is not present. A notch filter has been developed by experts to attenuate this RFI so that astronomical measurements can be taken in the day as well. As part of this work, we have also tried to analyse the response of the sample filter given to us.

\section{Experimental Setup}
In this work, we perform several experiments to study the performance of the dual-polarised bowtie antenna.
    \subsection{Studying the Polarisation of the EM waves}
    The dual-polarised bowtie antenna is placed on a stand and kept parallel to the ground. The elevation of the antenna is maintained using a spirit level. A dipole antenna is placed along the boresight of the bowtie antenna such that it is aligned parallel to one of the polarisation pairs. A TinySA is connected to the dipole and is used as a frequency source. The TinySA provides a configurable RF signal-generator output over a wide frequency range, allowing the required test frequency to be selected for the experiment. The length of the dipole is varied according to the frequency provided by the TinySA so that the dipole is resonant at that frequency. Hence, linearly polarised waves are emitted by the dipole and the dual-polarised bowtie is used to receive them by connecting it to an RTL-SDR \cite{stewart2015low}. RTL-SDR receivers can acquire and digitise RF signals and provide a low-cost software-defined-radio platform for experimental measurements. The dual-polarised bowtie is rotated a full circle in steps of 15° and the power values received by both orthogonal elements are measured at each step using GQRX software, keeping the RTL-SDR AGC off. GQRX is an open-source software-defined-radio receiver based on GNU Radio and Qt and supports RTL-SDR hardware; it can also be operated as an FFT-based measurement instrument for observing the received spectrum~\cite{gqrx}.

\begin{figure}[H]

 \caption{Experimental setup for polarisation-rotation observations}\label{ckt_diag}
 \includegraphics[width=10.5cm,height=8.5cm]{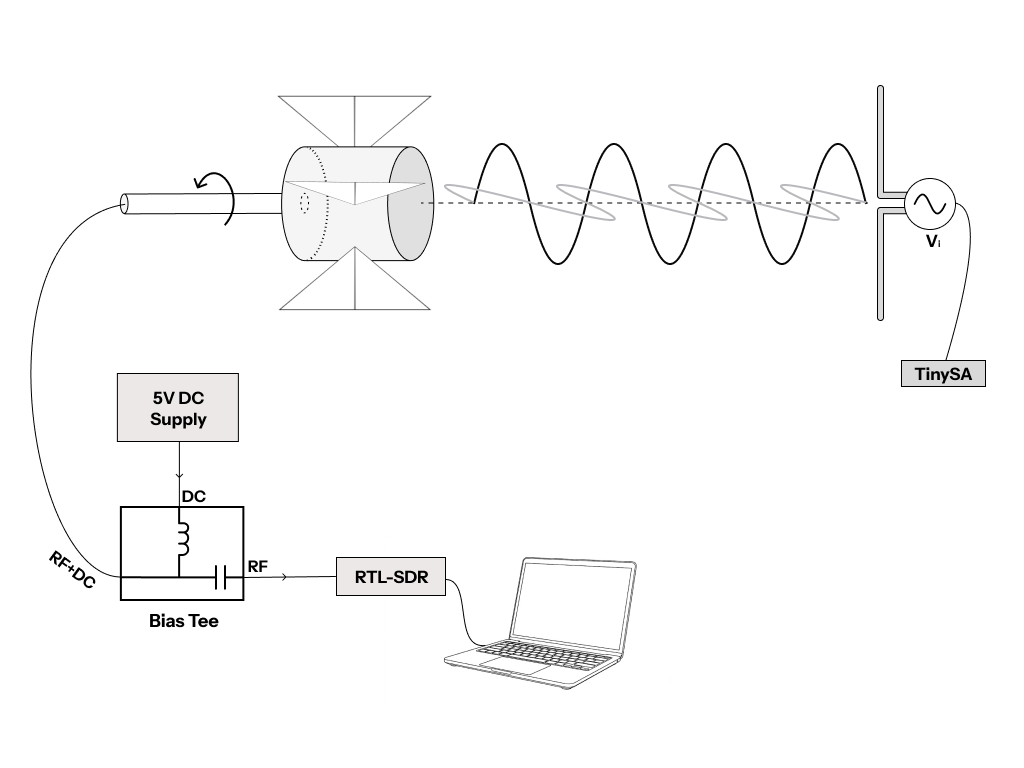}
 \end{figure}

\begin{figure}[H]
 \caption{The dual-polarised bowtie antenna is a) placed on a stand with its axis parallel to the ground b) with a disc attached to its top with angles marked in intervals of 15°. A nail protrudes from the top through the centre of the disc and a weight is hung from it so that the rotation angle can be measured at all times.}\label{bowtie}

 \begin{minipage}{0.48\textwidth}
  \centering
  \includegraphics[width=\textwidth]{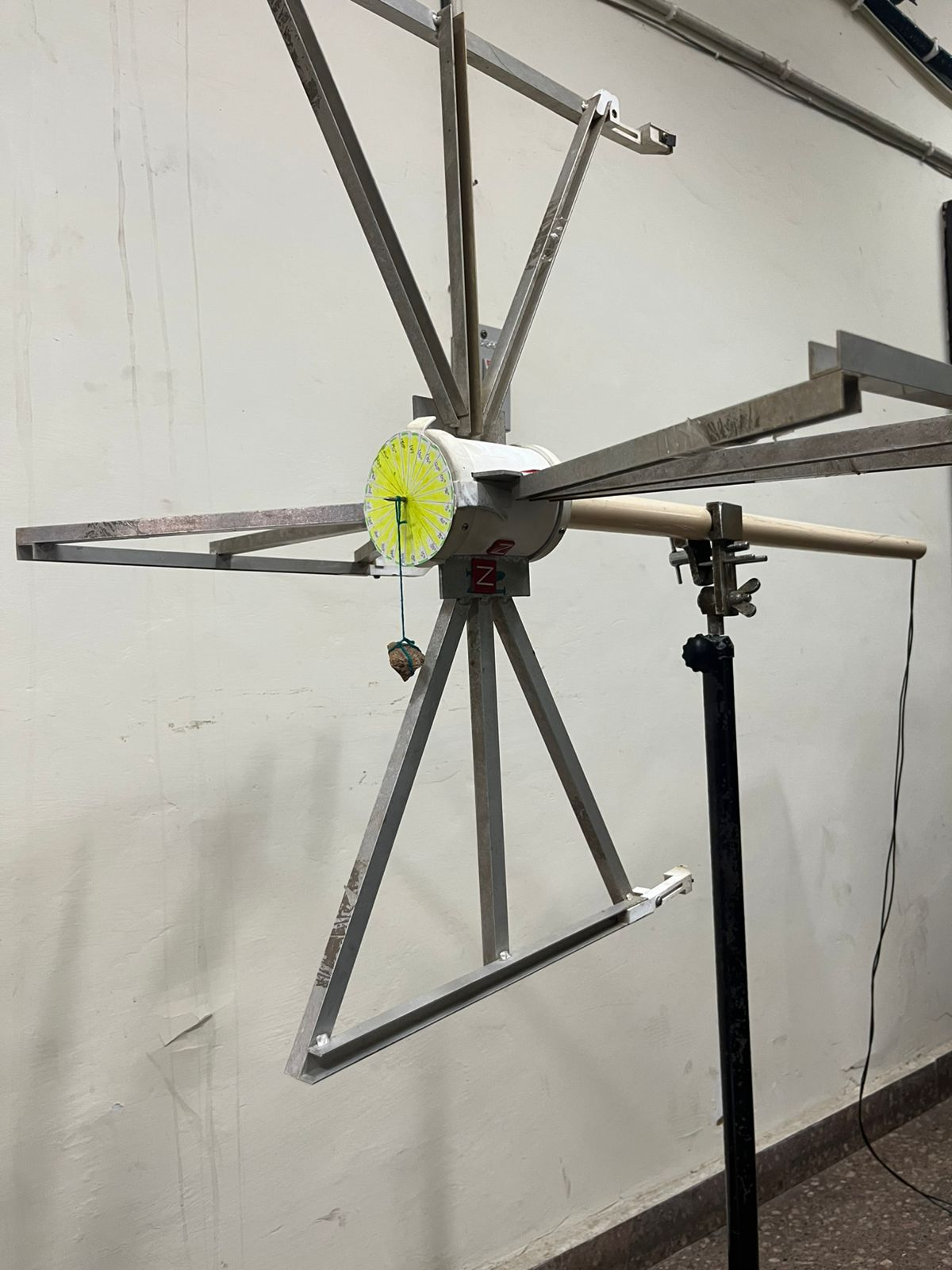}
  \\ (a)
 \end{minipage}
 \hfill
 \begin{minipage}{0.48\textwidth}
  \centering
  \includegraphics[width=\textwidth]{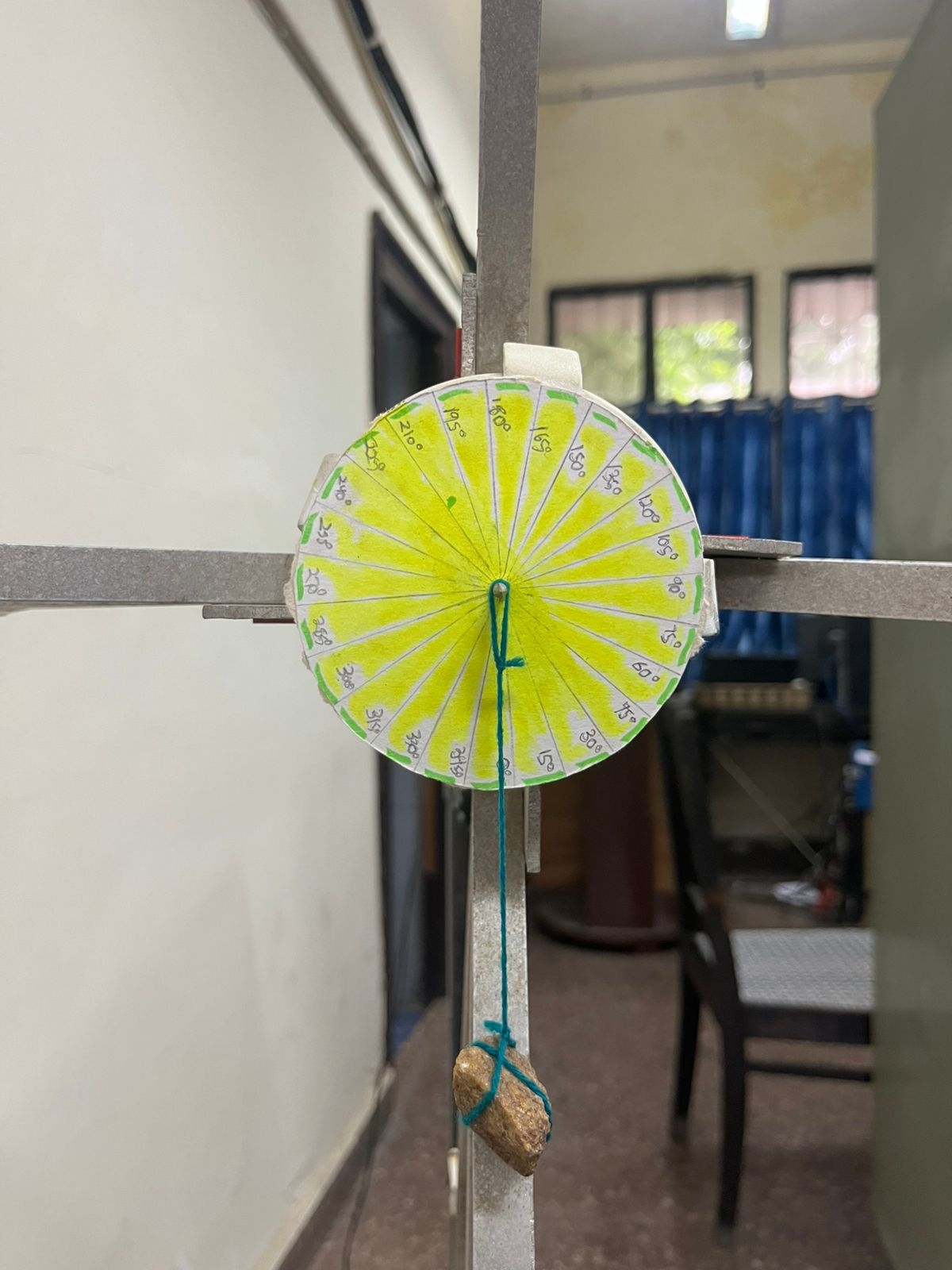}
  \\ (b)
 \end{minipage}

 \end{figure}
    
    \subsection{Response of the Notch Filter}
    We identify the operating frequency of the notch filter by taking its S-parameter measurements using a NanoVNA~\cite{derickson2021nanovna}. We then observe the frequency response of each polarisation of the antenna with and without the notch filter to find out the attenuation \cite{pozar2005microwave}.

 \begin{figure}[H]

 \caption{Experimental setup for observing the frequency response and attenuation of the bowtie antenna with the notch filter}\label{filter_response_ckt}
 \includegraphics[width=10.5cm,height=6.5cm]{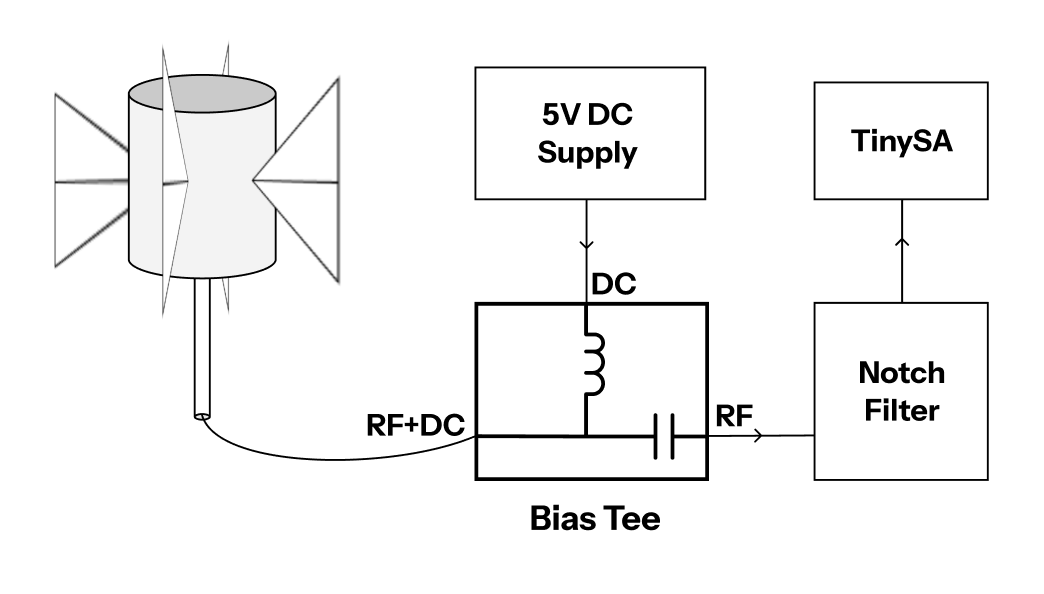}
 \end{figure}

\section{Method}
	  \subsection{Studying the Polarisation of the EM waves}
  Once the power values are measured for each rotation of the receiver antenna, the Malus equation is fitted to the data points to find  the polarisation angle of the source:
    \begin{equation}
        P = A\cos^2(\theta - \phi) + B \label{eq:malus}
    \end{equation}
    where $P$ is the measured power, $A$ is the power of the incident wave, $\theta$ specifies the angular position of the receiver antenna, $\phi$ is the angular position of the RF signal source and $B$ accounts for the baseline noise or stray radiation picked up by the receiver. The fitting is done for power in linear scale. The Malus equation fit predicts the best $\phi$ for a given set of data points. The axis of the source dipole was aligned parallel to the polarisation 1 bowtie arms. This orientation of the dual-pol bowtie is taken as $0^{\circ}$. The output power of the RF signal transmitted by the TinySA is fixed at $-27$~dBm. The expected value of $\phi $ is $0^\circ$ for polarisation 1 and $90^\circ$ for polarisation 2. To bring both estimates into a common reference frame, the fitted $\phi$ from Polarisation 2 is corrected by subtracting the $90^\circ$ offset before averaging. The mean of $\phi$ (from Polarisation 1) and the offset-corrected $\phi$ (from Polarisation 2) is then taken as the angular position of the source. \par
    This process is repeated for different frequencies of the source signal. The power measurements obtained across different rotation angles of the receiver are expected to carve a figure-of-eight on a polar graph, since the received power from a linearly polarised source varies as $\cos^2(\theta-\phi)$ for each port. The next step is to calculate cross-polarisation discrimination (XPD) at different frequencies. If $\beta$ is the best-fit angle for polarisation 1 at a specific frequency, and $P_{1\beta}$ and $P_{2\beta}$ are the power values of polarisation 1 and 2 at $\beta$, the cross-polarisation discrimination, in dB, for polarisation 1 is given as: 
    \begin{equation}
        10\log_{10}({P_{1\beta}/P_{2\beta}})
    \end{equation}
    Similarly if $\gamma$ is the best-fit angle for polarisation 2, then its cross-polarisation discrimination in dB is given by:
    \begin{equation}
        10\log_{10}{(P_{2\gamma}/P_{1\gamma})}
    \end{equation}
    However, the gain of both polarisation bowtie pairs is different. Hence, the power values of each polarisation are first normalised by dividing them by their respective maximum power values. Hence, the normalised values for polarisation 1 are:
    \begin{equation}
        P_{1norm} = P_1/P_{1\beta}
    \end{equation}
    and for polarisation 2 are:
    \begin{equation}
        P_{2norm} = P_2/P_{2\gamma}
    \end{equation}
    Hence, effectively, the cross-polarisation discrimination for polarisation 1 is:
    \begin{equation}
        10\log_{10}{(P_{2\gamma}/P_{2\beta})}
    \end{equation}
    and that for polarisation 2 is:
    \begin{equation}
        10\log_{10}{(P_{1\beta}/P_{1\gamma})}
    \end{equation}
Since there is a big difference in the obtained peak and null power values, an unweighted least-squares fit is dominated by the peak. An ordinary least-squares fit minimises the sum of squared absolute errors and hence a small fractional error at the peak contributes far more to the total error than a large fractional error at the null. This implies that peak accuracy is favoured over that of null, which affects the XPD estimate. This is corrected by performing the fit with weighted least squares, solved iteratively (Iteratively Reweighted Least Squares, IRLS)~\cite{green1984iteratively}, wherein the weight is given by:

\begin{equation}
    w_i = \frac{1}{P_{\mathrm{model},i}^{\,2\alpha}}
    \label{eq:irls_weight}
\end{equation}
Here, $P_{model,i}$ is the power predicted by the fit at the i-th rotation of the receiver antenna and $\alpha$ is a weighting exponent. Using the generalised least-squares (Aitken) theorem~\cite{aitken1936iv}, the best estimator under heteroscedastic noise is obtained by weighting each point by the inverse of its noise variance, $w_i = 1/\sigma_i^2$~\cite{bevington2003data}. Assuming that the standard deviation varies as $\sigma_i \propto P^{\alpha}_{model,i}$, we obtain the weight formula above. We assume $\alpha = 1$, which means that the noise is assumed to vary linearly with signal level or equivalently, to be uniform in dB-scale. Hence the weights are given by:
\begin{equation}
    w_i = 1/P^2_{model,i}
\end{equation}

    \subsection{Response of the Notch Filter} The notch filter is connected between the CH0 and CH1 ports of the NanoVNA. The notch filter is symmetric in nature, so flipping the connections of the filter produces no change in the measurements. The NanoVNA is then connected to a laptop running NanoVNASaver software, which is a GUI application used to observe and save the measurements taken using the NanoVNA~\cite{nanovnasaver}. The CH0 port measures the \(S_{11}\) parameter, which describes the reflection from the filter at a particular frequency and is expressed in terms of the incident and reflected powers as

\begin{equation}
S_{11}(\mathrm{dB}) =
10\log_{10}\left(
\frac{P_{\mathrm{reflected}}}{P_{\mathrm{incident}}} 
\right) \label{eq:s11}
\end{equation}

The CH1 port measures the \(S_{21}\) parameter, which represents the forward transmission through the filter and is expressed in terms of the transmitted and incident powers as

\begin{equation}
S_{21}(\mathrm{dB}) =
10\log_{10}\left(
\frac{P_{\mathrm{transmitted}}}{P_{\mathrm{incident}}} 
\right) \label{eq:s21}
\end{equation}

By plotting the S-parameters vs frequency we can find out the operating frequency of the notch filter. The notch filter is then connected to the bowtie antenna to observe its frequency response. Attenuation of the RFI can be found out by comparing the frequency response of the antenna with and without the notch filter.

\section{Results}
\subsection*{Polarisation response of the bowtie}
By measuring the power received at various angles between the axis of the
transmitting dipole and the receiving dual-polarised bowtie, we obtain the
polarisation response of each port. As the bowtie is rotated about its
boresight, the direction of arrival of the signal relative to the antenna, and
hence the antenna gain in that direction, remains fixed; only the alignment
between the incident electric field and the polarisation axis of each port
changes. The received power is therefore expected to follow the polarisation
mismatch factor $\cos^2\Delta\theta$, which is the form of
Equation~\eqref{eq:malus}. We fit this equation to the data points to predict
the angle of polarisation of the source, and calculate the $R^2$ value to see
how well the data follow this model. The results are shown in
Figure~\ref{polar_results}.
\begin{figure}[H]
 \centering
\caption{Received power (dBFS) vs rotation angle (°) polar plot, showing the
	polarisation response of the bowtie at 5 different frequencies within its
	operating range of 80 MHz to 320 MHz. The curves for the two polarisations are
	plotted in blue (polarisation 1) and red (polarisation 2).}\label{polar_results}

 \begin{minipage}{0.6\textwidth}
  \centering
  \includegraphics[width=\textwidth,height=5.7cm]{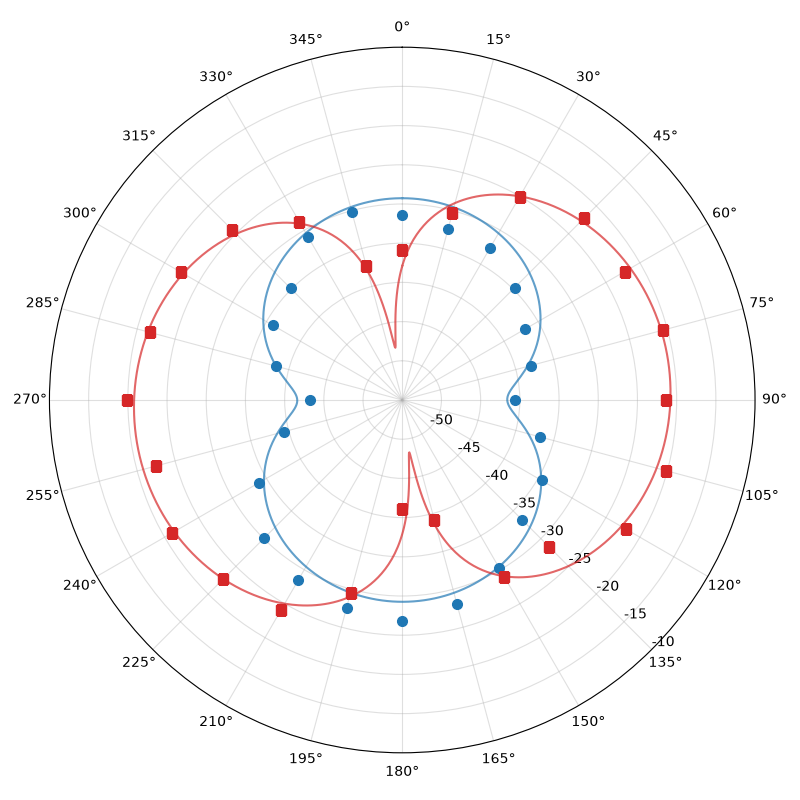}
  \\ (a) 135 MHz
 \end{minipage}
  \begin{minipage}{0.6\textwidth}
  \centering
  \includegraphics[width=\textwidth,height=5.7cm]{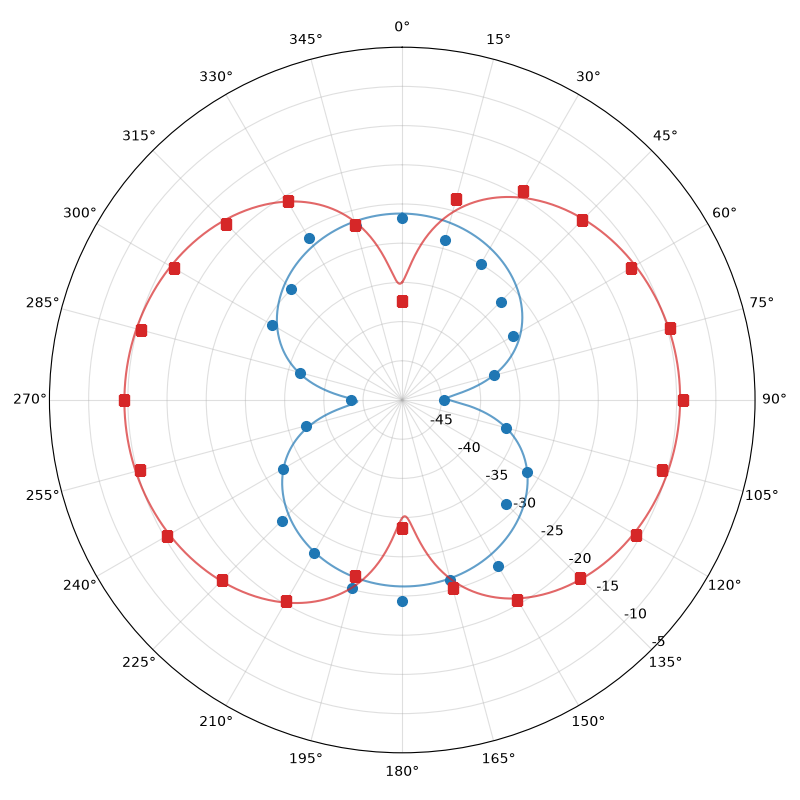}
  \\ (b) 150 MHz
 \end{minipage}
\end{figure}

\begin{figure}[H]
 \ContinuedFloat
 \centering
 \caption{(continued)}
 \begin{minipage}{0.6\textwidth}
  \centering
  \includegraphics[width=\textwidth,height=5.7cm]{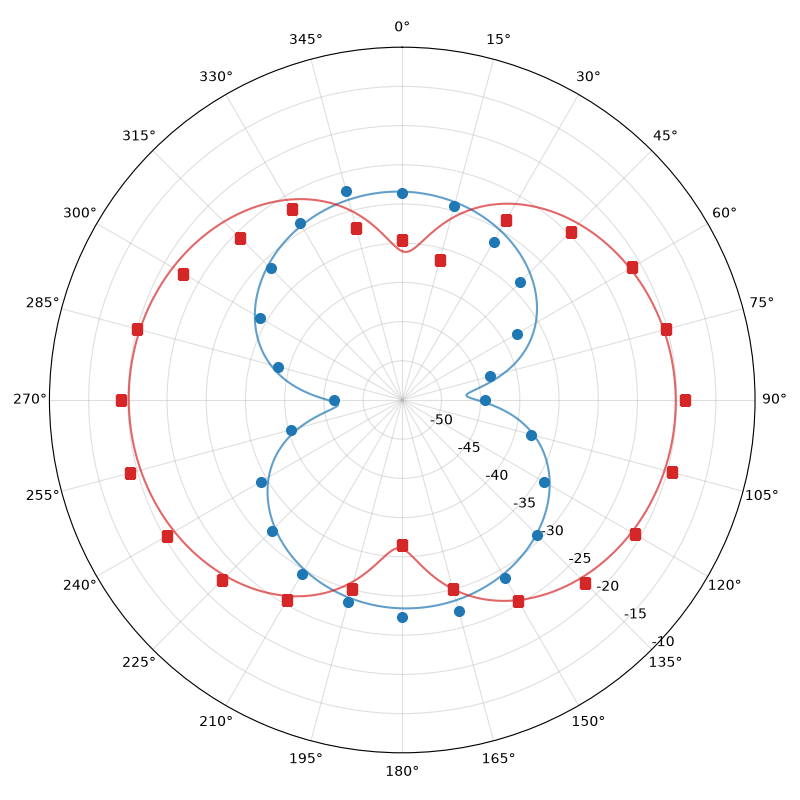}
  \\ (c) 258.62 MHz
 \end{minipage}
 \begin{minipage}{0.6\textwidth}
  \centering
  \includegraphics[width=\textwidth,height=5.7cm]{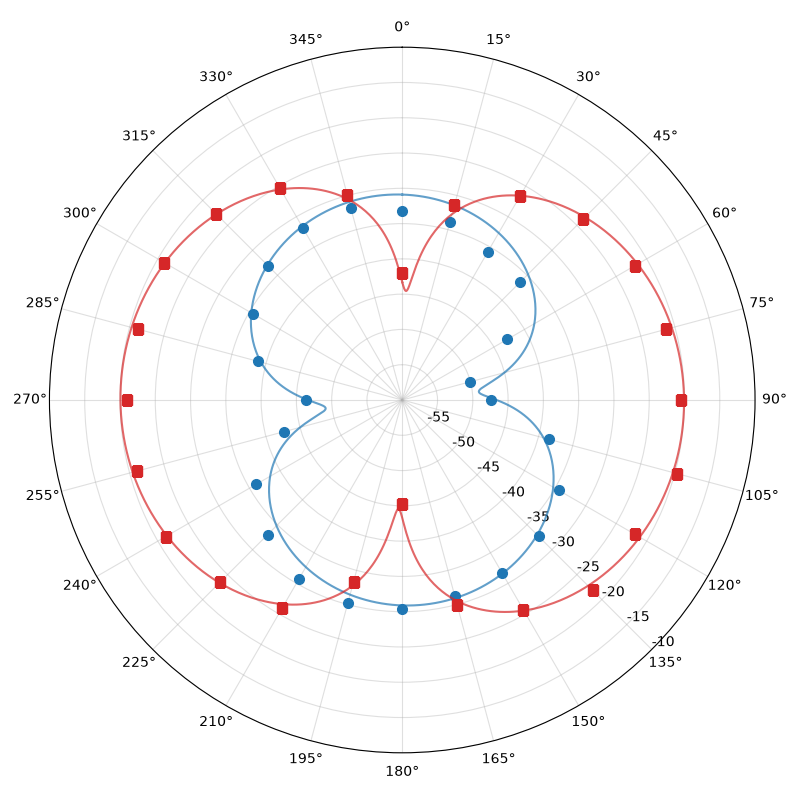}
  \\ (d) 270 MHz
 \end{minipage}
 \begin{minipage}{0.6\textwidth}
  \centering
  \includegraphics[width=\textwidth,height=5.7cm]{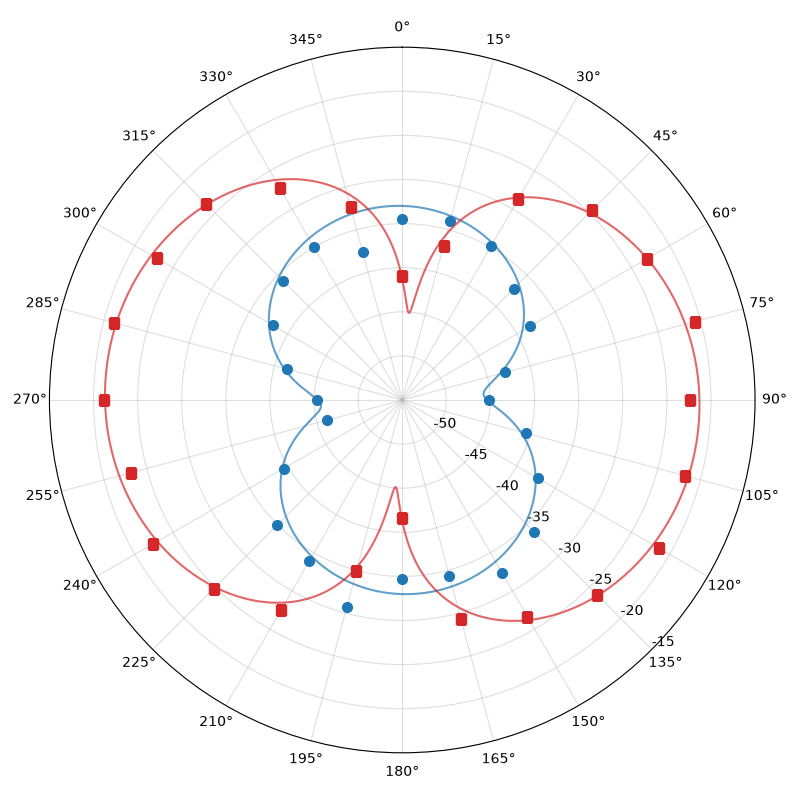}
  \\ (e) 308 MHz
 \end{minipage}
\end{figure}

\begin{table}
	\vspace{4pt}
	\caption{Predicted angle of polarisation of the source from the best fit of polarisation 1 ($\phi_1$), of polarisation 2 ($\phi_2$) and their mean ($\phi_{avg}$). Uncertainties are $\pm1\sigma$ statistical errors from the fit covariance.}
	\label{tab:angles}
	\renewcommand{\arraystretch}{1}
	\resizebox{\columnwidth}{!}{%
		\begin{tabular}{ccccc}
			\hline
			\bfseries S.no. & \bfseries Frequency (MHz) & \bfseries $\boldsymbol{\phi_{1}\,(^\circ)}$  & \bfseries $\boldsymbol{\phi_{2}\,(^\circ)}$  & \bfseries $\boldsymbol{\phi_{avg}\,(^\circ)}$ \\
			\hline
			1 & 135.00 & $-0.2 \pm 2.2$ & $-7.6 \pm 0.5$ & $-3.9 \pm 1.1$ \\
			2 & 150.00 & $-1.9 \pm 1.1$ & $-1.2 \pm 0.8$ & $-1.5 \pm 0.7$ \\
			3 & 258.62 & $-4.1 \pm 0.9$ & $1.3 \pm 1.4$ & $-1.4 \pm 0.8$ \\
			4 & 270.00 & $-6.2 \pm 1.2$ & $2.0 \pm 0.6$ & $-2.1 \pm 0.7$ \\
			5 & 308.00 & $-4.0 \pm 1.7$ & $4.4 \pm 0.5$ & $0.2 \pm 0.9$ \\
			\hline
		\end{tabular}%
	}
\end{table}

It is observed that the two curves are rotated by approximately 90° with
respect to each other, suggesting that the two polarisations are indeed
orthogonal. The eight-shaped figure arises because the received power is the
product of the antenna's co-polar response in the (fixed) direction of the
source and the polarisation mismatch factor $\cos^2(\Delta\theta)$. The measurements therefore confirm that each port is linearly polarised and that the two ports are orthogonal.
\subsection*{Polarisation and Discrimination}
By fitting Equation~\eqref{eq:malus} to observations such as those in Figure~\ref{polar_results},  we can find the angle of the incident linearly polarised signal with respect to the chosen reference. In Table~\ref{tab:angles}, $\phi_{avg}$ is the predicted angle of the incident linearly polarised signal, which was manually set at $0^\circ$. $\phi_{avg}$ deviates from the true angle by less than $5^\circ$ at all five frequencies, although the individual fits deviate by up to $7.6^\circ$ (polarisation 2 at 135 MHz). The quoted uncertainties are statistical only; the manual alignment of the transmitting dipole and the $15^\circ$ markings on the rotation disc introduce additional systematic errors that are not included in them and account for these larger offsets.

\begin{figure}[H]
\centering
\caption{Cross-polarisation Discrimination (dB) vs Frequency (MHz) for polarisation 1 (blue circles) and polarisation 2 (red squares)}
\includegraphics[scale=0.45]{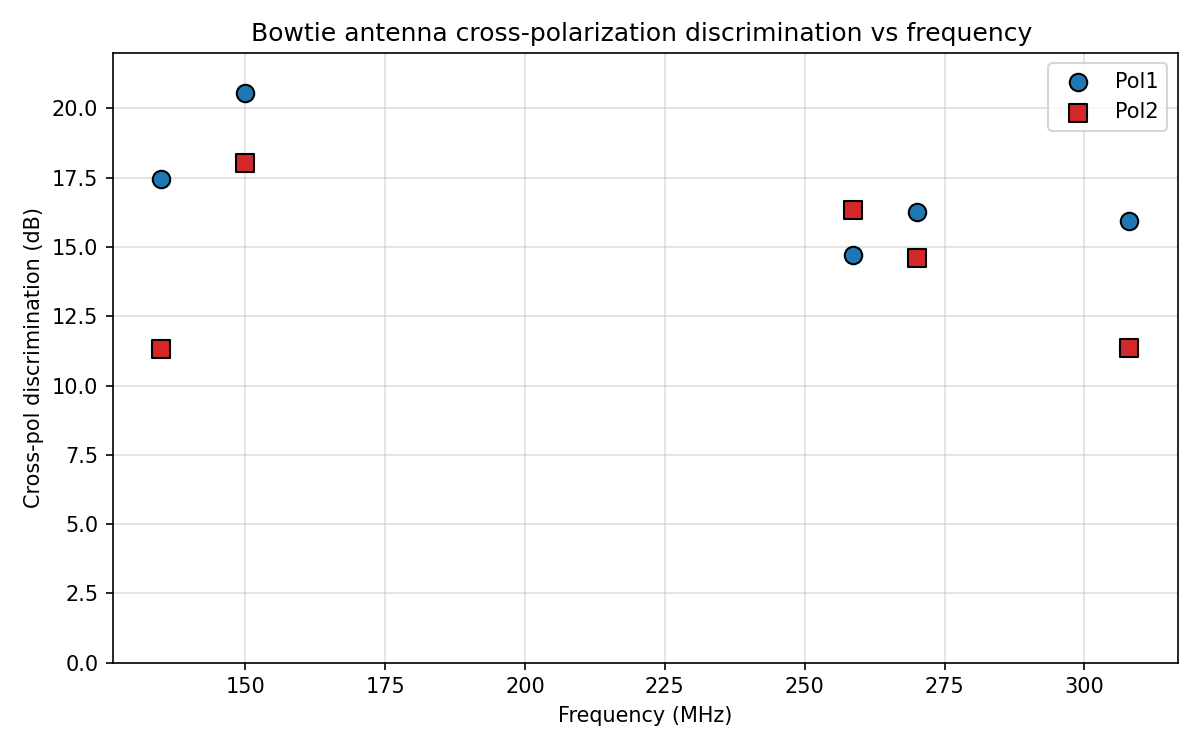} \label{fig:xpd}

\end{figure}

\begin{table}[H]
	\centering
	\caption{Cross-polarisation discrimination (XPD) and $R^2$ fit quality at each frequency. XPD uncertainties are $\pm1\sigma$ statistical errors from the fit covariance.}
	\label{tab:discrimination}
	\vspace{4pt}
	\renewcommand{\arraystretch}{1.4}
	\resizebox{\columnwidth}{!}{%
		\begin{tabular}{cccccc}
			\hline
			\bfseries S.no. & \bfseries Frequency (MHz) & \bfseries XPD$\boldsymbol{_{pol1}}$ (dB) & \bfseries XPD$\boldsymbol{_{pol2}}$ (dB) & \bfseries $\boldsymbol{R^2}\boldsymbol{_{pol1}}$ & \bfseries $\boldsymbol{R^2}\boldsymbol{_{pol2}}$ \\
			\hline
			1 & 135.00 & $17.4 \pm 2.5$ & $11.3 \pm 1.1$ & 0.84 & 0.96 \\
			2 & 150.00 & $20.5 \pm 0.9$ & $18.0 \pm 1.0$ & 0.94 & 0.98 \\
			3 & 258.62 & $14.7 \pm 1.0$ & $16.3 \pm 1.1$ & 0.96 & 0.91 \\
			4 & 270.00 & $16.3 \pm 1.2$ & $14.6 \pm 1.1$ & 0.88 & 0.99 \\
			5 & 308.00 & $15.9 \pm 1.5$ & $11.4 \pm 0.9$ & 0.87 & 0.98 \\
			\hline
		\end{tabular}%
	}
\end{table}
 The cross-polarisation discrimination (XPD) is measured and is used to understand the leakage between the two polarisations. Since XPD of polarisation 2 is lower than polarisation 1 in most of the sampled frequencies, there is more leakage from polarisation 2 into polarisation 1 than the other way. Figure~\ref{fig:xpd} plots the XPD of the bowtie vs frequency to show that the XPD of both polarisations are around or less than 20 dB, which is on the lower side.

\subsection*{Filter response}
\begin{figure}[htbp]

 \caption{$S_{11}$ and $S_{21}$ magnitudes (dB) vs Frequency (MHz) for the notch filter}\label{Filterresponse}
 \includegraphics[width=10cm,height=4.8cm]{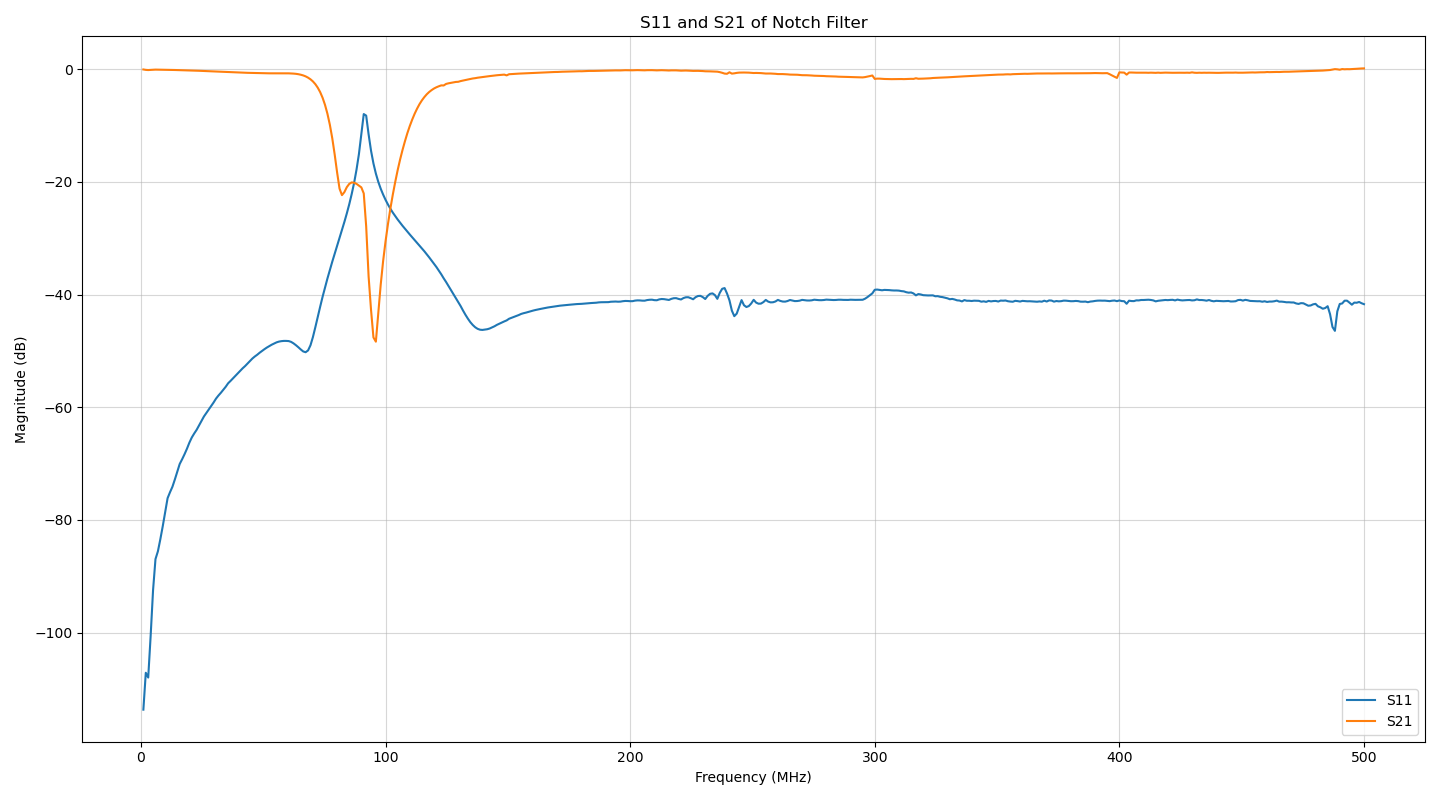}
 \end{figure}

As seen in Figure~\ref{Filterresponse}, the $S_{11}$ magnitude shows a steep rise of more than 40 dB around 100 MHz. The trend shows that the notch filter transmits all signals except for a band around 100 MHz. $S_{21}$ magnitude also drops around the same range implying that signals in this range are filtered out. Therefore, the notch filter attenuates signals in the range of 80--120~MHz and can be used to suppress RFI in this range. The $-3$~dB stopband extends from about 72 to 121~MHz. The rejection exceeds 20~dB from about 81 to 103~MHz, with a minimum of about $-48$~dB near 96~MHz, and falls to about 12~dB at the upper edge of the FM band (108~MHz).

\begin{figure}[htbp]
\centering
\caption{RFI survey of bowtie antenna for both polarisations with and without notch filter. Attenuation values are high in the range 80--120~MHz while fluctuating around 0 dB for other frequencies, implying a good filter response and strong attenuation of RFI due to FM stations}
\includegraphics[scale=0.39]{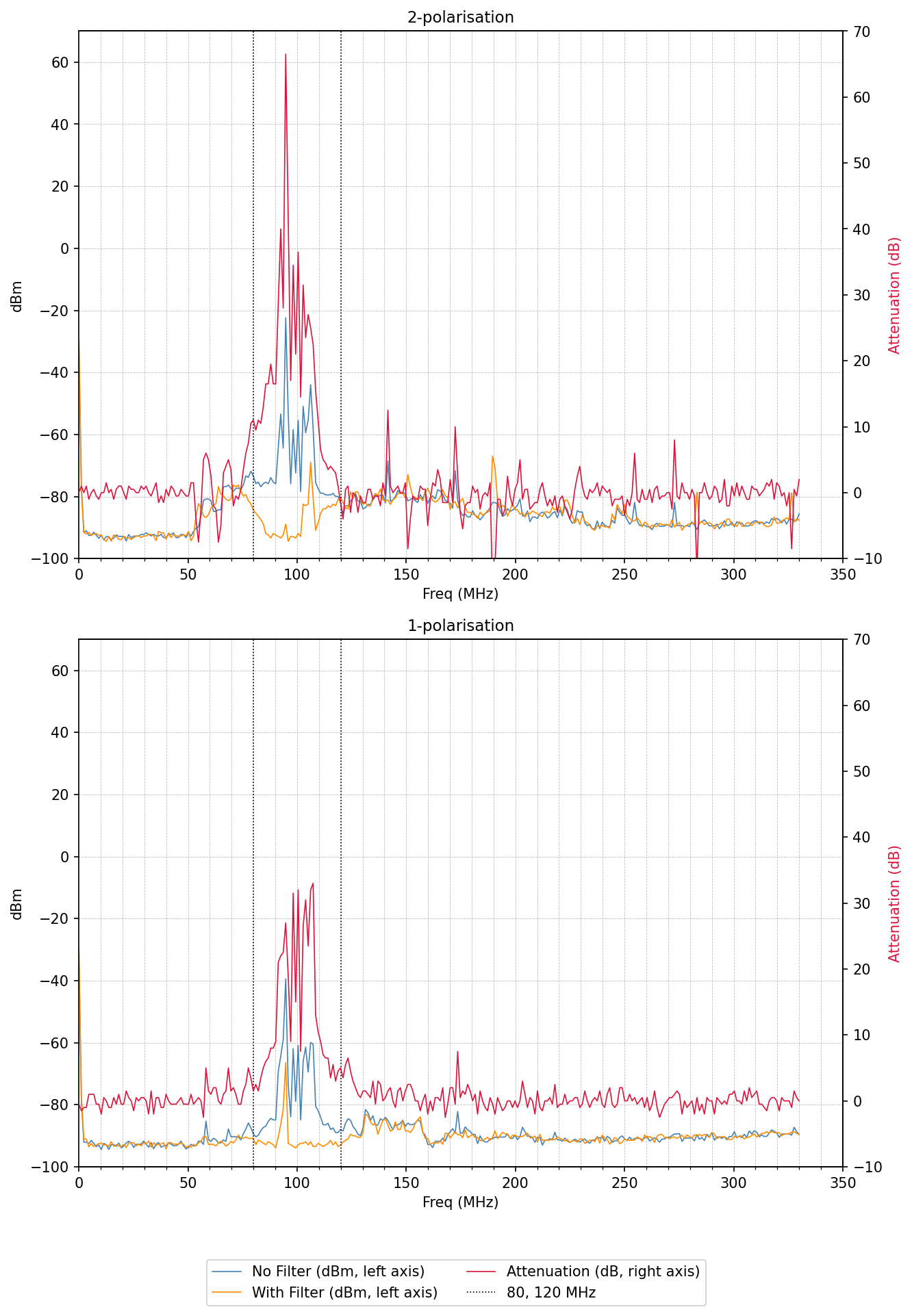}\label{fig:attenuation}
\end{figure}

The attenuation is high in the range of 80--120~MHz, as seen in Figure~\ref{fig:attenuation}, while the neighbouring region outside the band remains unaffected at the noise floor level. For both polarisations, the RFI due to FM stations of Delhi is suppressed strongly. Therefore, measurements can be taken during the day as well using the SWAN setup with the help of the notch filter. 

\section{Discussion, Limitations and Future Prospects}

The polarisation-rotation measurements confirm that the dual-polarised bowtie antenna responds to incident linear polarisation in a manner consistent with the Malus-law model, with high-quality fits ($R^2$ generally above 0.85, and often above 0.95) across the tested frequency range. The recovered source angles $\phi_{avg}$ deviate from the true $0^\circ$ orientation by less than $5^\circ$ at all five frequencies, which is a reasonable level of agreement given the manual alignment of the transmitting dipole and the bowtie antenna. \par The measured XPD values, hovering around or below 20 dB, indicate a moderate and non-negligible amount of leakage between the two polarisation channels, and since the XPD of polarisation 2 is lower than that of polarisation 1 at four of the five frequencies, this leakage appears to be asymmetric, with more power from polarisation 2 spilling into polarisation 1 than vice versa; this asymmetry could arise from small differences in the electrical and mechanical construction of the two bowtie arm pairs. \par The notch filter characterisation shows a well-defined stopband centred close to 100 MHz, with $S_{21}$ dropping by more than 40 dB and $S_{11}$ rising sharply in the same range. The resulting RFI attenuation strongly suppresses the Delhi FM band for both polarisations, although stations near the upper edge of the band (above about 103~MHz) are attenuated less, enabling daytime SWAN observations.

A few limitations of the present study should be noted: the rotation measurements were carried out manually in $15^\circ$ steps using a disc-and-weight arrangement to track the antenna angle, which limits the angular resolution and introduces some uncertainty in the exact rotation angle at each step. The experiments were performed in an uncontrolled environment rather than an anechoic chamber, so reflections from the floor and walls, nearby scatterers, and ambient RFI could all contribute to the measured polarisation response and reduce the fit quality at some frequencies. The frequency coverage of the current measurements, five discrete frequencies between 135 and 308 MHz, is sparse compared to the full 80--320~MHz operating band of the antenna, so finer frequency sampling would be needed to fully characterise the polarisation response. 

Building on this work, future measurements could use a 3D printed rotation-and-measurement setup using a stepper motor and microcontroller like Arduino, allowing automatic, finer angular sampling and removing possible human error. With this setup, several full cycles of rotations can be averaged to give a much more accurate dataset and also significantly reduce deviation in $\phi$ from the true value. This technique can be used to check whether the signal from a non-variable radio source in space is linearly polarised, since the received power from a linearly polarised signal varies as $\cos^2$ of the angle between the signal
polarisation and the antenna axis, producing a distinct eight-shaped figure on
rotation. This technique may be used to estimate the Faraday rotation measure (RM) along the line of sight, since the polarisation angle $\psi$ of a linearly polarised wave passing through the Earth's ionosphere varies as
\begin{equation}
    \psi(\lambda^2) = \psi_0 + \mathrm{RM}\,\lambda^2 \label{eq:faraday}
\end{equation}
where $\psi_0$ is the intrinsic polarisation angle at $\lambda \to 0$, and the rotation measure is given by
\begin{equation}
    \mathrm{RM} = \frac{e^3}{2\pi m_e^2 c^4}\int_0^L n_e\, B_\parallel \, dl \label{eq:rm}
\end{equation}
Here, $e$ and $m_e$ are the electron charge and mass, $c$ is the speed of light, $n_e$ is the free-electron density, $B_\parallel$ is the line-of-sight component of the magnetic field, and the integral is taken along the propagation path length $L$ through the ionosphere. By measuring $\psi$ at several frequencies and fitting Equation \eqref{eq:faraday}, the slope of $\psi$ vs $\lambda^2$ directly gives the ionospheric RM \cite{rybicki1979radiative,brentjens2005faraday}. The intercept will give the intrinsic polarisation angle at the source. \par
A denser frequency sweep across the full operating band, ideally automated using the TinySA and RTL-SDR in a scripted acquisition loop, would allow the frequency dependence of XPD and beamwidth to be mapped more completely~\cite{stewart2015low}. Since this antenna forms a single element of the SWAN array, extending the present single-antenna characterisation to multiple deployed elements would help quantify element-to-element variation~\cite{tingay2013murchison,rao2021wide}. 
The experiments were performed and the data were analysed in the undergraduate laboratory of St.~Stephen's College and helped students visualise the concept of polarisation and filters.

\section*{Acknowledgements}
The authors are grateful to Prof. Avinash Deshpande for his constant support and to the Indian Institute of Technology, Kanpur for providing the notch filter. The authors are also grateful to Dr. Jayasandhya Meenakshinathan for discussions and her guidance. The authors are grateful to the Principal, St.~Stephen's College for her encouragement and support by
providing the facility to conduct experiments.

\bibliographystyle{unsrt}
\bibliography{citte}

\end{document}